# Demkov–Kunike Model for Cold Atom Association: Weak Interaction Regime

**R. S. Sokhoyan**[a,b,*], **H. H. Azizbekyan**[a,c,d], **C. Leroy**[b], **and A. M. Ishkhanyan**[a]

*[a]Institute for Physical Research, NAS of Armenia, Ashtarak, Armenia*

*[b]Institut Carnot de Bourgogne, Université de Bourgogne, Dijon, France*

*[c]Laboratoire de Physique Moléculaire et des Collisions, Université Paul Verlaine – Metz, Metz Cedex 3, France*

*[d]Moscow Institute of Physics and Technology, Dolgoprudny, Russia*

**Abstract**—We study the nonlinear mean-field dynamics of molecule formation at coherent photo- and magneto-association of an atomic Bose–Einstein condensate for the case when the external field configuration is defined by the quasi-linear level crossing Demkov–Kunike model, characterized by a bell-shaped pulse and finite variation of the detuning. We present a general approach to construct an approximation describing the temporal dynamics of the molecule formation in the weak interaction regime and apply the developed method to the nonlinear Demkov–Kunike problem. The presented approximation, written as a scaled solution to the linear problem associated to the nonlinear one we treat, contains fitting parameters which are determined through a variational procedure. Assuming that the parameters involved in the solution of the linear problem are not modified, we suggest an analytical expression for the scaling parameter.

## 1. INTRODUCTION

Physics of ultracold trapped gases has become a rapidly developing research domain due to recent experimental and theoretical achievements (see reviews [1–3]). One of the interesting research directions in this field is coherent molecule formation in atomic quantum gases via application of associating optical or magnetic fields (such processes are referred to as "superchemistry" [4]) which under certain experimental conditions [5, 6] can be described by a basic mean-field time-dependent two-level problem, defined by the following set of coupled nonlinear equations [7]:

$$i\left(da_1/dt\right) = U(t)e^{-i\delta(t)}\overline{a}_1 a_2,$$
$$i\left(da_2/dt\right) = \left(U(t)/2\right)e^{i\delta(t)}a_1 a_1, \tag{1}$$

where $t$ is time, $a_1$ and $a_2$ are the atomic and molecular state probability amplitudes, respectively, $\overline{a}_1$ denotes the complex conjugate of $a_1$, the real function $U(t)$ is referred to as the Rabi frequency of the associating field, and the real function $\delta(t)$ is the integral of the associated frequency detuning. System (1) conserves the total number of particles that we normalize to unity: $\left|a_1\right|^2 + 2\left|a_2\right|^2 = \text{const} = 1$. We will consider the condensate initially being in pure atomic state: $a_1(-\infty) = 1, a_2(-\infty) = 0$.

In the most of the theoretical developments (see, e.g., [8–12]) the dynamics of molecule formation has typically been treated by the constant-amplitude linear level-crossing Landau–Zener (LZ) model [13] (Fig. 1):

$$U = U_0, \quad \delta_t = 2\delta_0 t \tag{2}$$

(under conditions considered here the two techniques, photoassociation [14] and Feshbach resonance [15], are mathematically equivalent). However, the actual external field configuration applied in the experiments [16–18] is different from that defined by the LZ model. Hence, for the understanding of physics underlying these experiments, it is important to study how the variation of the pulse shape and the detuning affects the nonlinear dynamics of the system. The field configuration we discuss below is the first Demkov–Kunike (DK) quasi-linear level-crossing model [19], characterized by a bell-shaped pulse and a finite variation of the detuning (Fig. 1):

$$U = U_0 \text{sech}(t/\tau), \quad \delta_t = 2\delta_0 \tanh(t/\tau), \tag{3}$$

where $\tau$ is a positive parameter. The DK model is considered as a natural physical generalization of the LZ model. Without loss in generality we put $\tau = 1$ in what follows.

*E-mail: sruzan@gmail.com

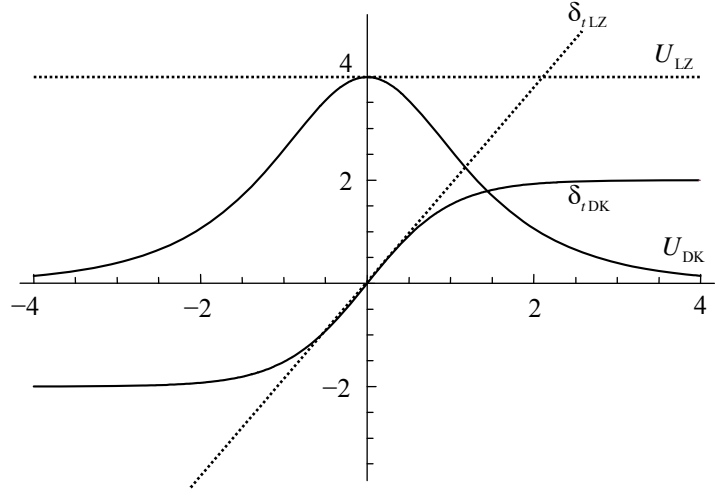

**Fig. 1.** Solid curves – the first Demkov–Kunike model: $U = U_0 \text{sech}(t)$, $\delta_t = 2\delta_0 \tanh(t)$; dotted lines – the Landau–Zener model: $U = U_0$, $\delta_t = 2\delta_0 t$.

The goal of the present paper is to study the weak interaction regime of the nonlinear DK problem (defined by Eqs. (1) and (3)). This regime describes the case when the number of molecules, formed during the association process, is small. For the DK model, it corresponds to the weak coupling limit ($U_0 \ll 1$, $\forall \delta_0$) and the very large detuning regime of the strong coupling limit ($U_0 > 1$, $\delta_0 \gg U_0$).

The weak interaction regime of the nonlinear DK problem has already been discussed in [20], where, using an exact nonlinear Volterra integral equation [21] and applying Picard's successive approximation, an analytic formula for the final probability of transition to the molecular state has been obtained. However, the whole temporal dynamics of the system has not been treated there. In the present paper we develop a method to describe the time evolution of the system in the mentioned regime.

## 2. MATHEMATICAL TREATMENT

The present development is based on the following exact equation for the molecular state probability $p = |a_2|^2$ [22]:

$$p_{ttt} - \left(\frac{\delta_{tt}}{\delta_t} + 2\frac{U_t}{U}\right)p_{tt} + \left[\delta_t^2 + 4U^2(1-3p) - \left(\frac{U_t}{U}\right)_t + \frac{U_t}{U}\left(\frac{\delta_{tt}}{\delta_t} + \frac{U_t}{U}\right)\right]p_t +$$
$$+ \frac{U^2}{2}\left(\frac{\delta_{tt}}{\delta_t} - \frac{U_t}{U}\right)(1 - 8p + 12p^2) = 0. \tag{4}$$

First we simplify this equation, applying transformation of the independent variable

$$z(t) = \int_0^t \frac{U(t')}{U_0} dt', \tag{5}$$

reducing Eq. (4) to the following constant-amplitude form:

$$p_{zzz} - \frac{\delta_{zz}^*}{\delta_z^*}p_{zz} + \left[\delta_z^{*2} + 4U_0^2(1-3p)\right]p_z + \frac{U_0^2}{2}\frac{\delta_{zz}^*}{\delta_z^*}\left(1 - 8p + 12p^2\right) = 0, \tag{6}$$

where the effective detuning $\delta_z^*$ is defined as

$$\delta_z^*(z(t)) = \delta_t(t)\big[U_0/U(t)\big] \,. \tag{7}$$

In the case of the DK model (3), relations (5) and (7) take the following form:

$$z(t) = 2\arctan(e^t) - \pi/2, \quad z \in (-\pi/2, \pi/2), \tag{8}$$

$$\delta_z^* = 2\delta_0\tan(z), \quad \delta_z^*(z(t)) = 2\delta_0\sinh(t) \,. \tag{9}$$

Note that the resonance crossing point $t = 0$ is mapped onto the point $z = 0$.

A linear problem associated to the nonlinear problem at hand can be defined by removing the nonlinear terms from equation for the molecular state probability (4). It can be checked that the obtained linear equation is obeyed by the function $p_L = |a_{2L}|^2$, where $a_{2L}$ is determined from the linear set of equations

$$\begin{aligned} i\,da_{1L}/dt &= U(t)e^{-i\delta(t)}a_{2L}, \\ i\,da_{2L}/dt &= U(t)e^{i\delta(t)}a_{1L}, \end{aligned} \tag{10}$$

with the following normalization constraint:

$$|a_{1L}|^2 + |a_{2L}|^2 = I_L = 1/4 \,. \tag{11}$$

From the quantum optics point of view, this system describes coherent interaction of an isolated atom with optical laser radiation [23].

Next, we write the exact solution of the linear set (10) satisfying the initial conditions $a_1(-\infty) = 1$ and $a_2(-\infty) = 0$, hence, normalized as

$$|a_{1DK}|^2 + |a_{2DK}|^2 = I_L = 1 \,. \tag{12}$$

This solution is given as follows [19]:

$$a_{1DK} = {}_2F_1(-i\delta_0 + \sqrt{U_0^2 - \delta_0^2}, -i\delta_0 - \sqrt{U_0^2 - \delta_0^2}; 1/2 - i\delta_0; x), \tag{13}$$

$$a_{2DK} = \frac{U_0}{i + 2\delta_0}(\cosh(t))^{-1+2i\delta_0} \, {}_2F_1(1 - i\delta_0 + \sqrt{U_0^2 - \delta_0^2}, 1 - i\delta_0 - \sqrt{U_0^2 - \delta_0^2}; 3/2 - i\delta_0; x), \tag{14}$$

where $x(t) = (1 + \tanh(t))/2$, and ${}_2F_1(\alpha; \beta; \gamma; x)$ is the Gauss hypergeometric function [24]. Accordingly, the probability of transition to the second level is written as

$$p_{DK} = |a_{2DK}|^2, \tag{15}$$

and the final transition probability is given by the formula

$$p_{DK}(+\infty) = |a_{2DK}(+\infty)|^2 = 1 - \cos^2\left(\pi\sqrt{U_0^2 - \delta_0^2}\right)\operatorname{sech}^2\pi\delta_0 \,. \tag{16}$$

It can easily be seen that the solution of set (10), normalized to 1/4 (according to the normalization condition (11)), is given as

$$a_{1L} = \frac{a_{1DK}}{2} \quad \text{and} \quad a_{2L} = \frac{a_{2DK}}{2} \,. \tag{17}$$

To develop better intuitive understanding of the problem at hand, we examine the exact equation for the molecular state probability (4). The nonlinearity is determined by the current value of the transition probability $p$. Hence, one may expect that if $p$ remains small enough (note that, anyway, $p \leq 1/2$) the role of the nonlinearity will be rather restricted. In this case, neglecting the nonlinear terms in Eq. (4), we get the linear equation, satisfied by the function $p_L = |a_{2L}|^2$ (see Eqs. (15) and (17)). Studying now the solution of the linear two-state problem $p_L(t)$ we see that, if the dimensionless peak Rabi frequency $U_0$ is small enough ($U_0 << 1$) or if it is much smaller as compared to the sweep rate through the resonance ($U_0 << \delta_0$) then the function $p_L$ does not attain large values. From this one can infer that in these cases the transition probability defined by the nonlinear two-state problem is close to that defined by the linear two-state problem. This observation suggests that in the weak interaction regime the temporal dynamics of the molecule formation could be described by the scaled solution of the linear problem, but with some effective parameters $U_0^*$ and $\delta_0^*$:

$$p_0 = C^* \left[ p_L(U_0^*, \delta_0^*, t) \big/ p_L(U_0^*, \delta_0^*, +\infty) \right]. \tag{18}$$

Such a conjecture for the weak coupling limit of the LZ model was made in [25] where an accurate analytic approximation written in terms of the scaled solution to an auxiliary linear LZ problem with some effective LZ parameter had been constructed and analytical expressions for the introduced parameters had been determined. Preliminary numerical analysis shows that function (18) is capable to provide high enough accuracy without modification of the detuning parameter $\delta_0$. Hence, hereafter we put $\delta_0^* = \delta_0$. Note that for the variational ansatz (18) the approximate expression for the final transition probability is defined by the value of the scaling parameter $C^*$ only.

To develop general principles from which the fitting parameters $U_0^*$ and $C^*$ could be determined, we insert the suggested ansatz $p_0$ into the transformed equation for the molecular state probability (6) and consider the behavior of the remainder

$$R = \left[ (d/dz) - \left( \delta_{zz}^* / \delta_z^* \right) \right] r(z), \tag{19}$$

where $r(z)$ is the notation for

$$r(z(t)) = C^* \frac{U_0^{*2}}{2} \left[ 4 - 8 \frac{p_{DK}(U_0^*, t)}{p_{DK}(U_0^*, +\infty)} \right] - \frac{U_0^2}{2} \left[ 1 - 8 C^* \frac{p_{DK}(U_0^*, t)}{p_{DK}(U_0^*, +\infty)} + 12 \left( C^* \frac{p_{DK}(U_0^*, t)}{p_{DK}(U_0^*, +\infty)} \right)^2 \right]. \tag{20}$$

It is intuitively understood that the better approximation $p_0$ is the smaller remainder $R$ will become (it would identically be zero if $p_0$ is the exact solution to Eq. (6))]. Thus, we try to minimize the remainder via appropriate choice of the fitting parameters $C^*$ and $U_0^*$. We first note that, since the function $p_{DK}(U_0^*, t)$ is bounded everywhere, the function $R$ is bounded almost everywhere. The exceptions are the resonance crossing point $z = 0$ $(t = 0)$ and the points $z = \pm \pi/2$ $(t = \pm \infty)$, where, due to the term $\delta_{zz}^* / \delta_z^*$ in the operator $(d/dz - \delta_{zz}^*/\delta_z^*)$, in general, $R$ diverges. Since when passing to the physical variable $t$ the singularities of the remainder $R$ at $z = \pm \pi/2$ disappear, we choose to eliminate divergence at the resonance crossing point $z = 0$, i.e., we require $U_0^*$ and $C^*$ to satisfy the equation $r(0) = 0$. Explicitly, this equation is written as

$$C^* \frac{U_0^{*2}}{2} \left[ 4 - 8 \frac{p_{DK}(U_0^*, 0)}{p_{DK}(U_0^*, +\infty)} \right] - \frac{U_0^2}{2} \left[ 1 - 8 C^* \frac{p_{DK}(U_0^*, 0)}{p_{DK}(U_0^*, +\infty)} + 12 \left( C^* \frac{p_{DK}(U_0^*, 0)}{p_{DK}(U_0^*, +\infty)} \right)^2 \right] = 0. \tag{21}$$

To find appropriate values for parameters $U_0^*$ and $C^*$, we need to introduce one more equation. Of course, in order to construct an approximation as simple as possible, one may first try to avoid variation of both of the auxiliary parameters and try to get a simpler, one-parametric approximation instead. A natural choice is then to fix $U_0^* = U_0$ and vary $C^*$ alone. Equation (21) then readily gives:

$$C^* = \lim_{t \to +\infty} p_0(t) = p_{DK}(U_0, +\infty) \left( 1 - \sqrt{1 - 3 p_{DK}(U_0, 0)^2} \right) \Big/ 6 p_{DK}(U_0, 0)^2. \tag{22}$$

As it follows from Eqs. (12) and (13), the explicit expression for $p_{DK}(0)$ can be written as:

$$p_{DK}(0) = 1 - \left| {}_2F_1\left( -i\delta_0 + \sqrt{U_0^2 - \delta_0^2}, -i\delta_0 - \sqrt{U_0^2 - \delta_0^2}; 1/2 - i\delta_0; 1/2 \right) \right|^2. \tag{23}$$

Numerical analysis shows that the constructed approximation (18), with $\delta_0^* = \delta_0$, $U_0^* = U_0$ and $C^*$ defined according to Eq. (22), accurately describes the temporal dynamics of the molecule formation in the weak coupling limit. Further, we compare the derived approximate expression for the final transition probability (22) with that calculated in [20]:

$$\lim_{t \to +\infty} p(t) \approx \frac{p_{DK}(U_0, +\infty)}{4} \left( 1 + \frac{3 U_0^2}{64} \frac{1 + 2\delta_0^2}{1 + \delta_0^2} \left| B(1/2 + i\delta_0, 1/2 + i\delta_0) \right|^2 p_{DK}(U_0, +\infty) \right). \tag{24}$$

The derived formula (22) for the final transition probability, approximation (24) of [20], and the result of numerical simulation are shown in Fig. 2. As we see, in the weak coupling limit $U_0 < 1$ the derived formula works slightly better than the one defined by Eq. (24). On the other hand, numerical analysis shows that in the very large detuning regime of the strong coupling limit ($U_0 > 1$, $\delta_0 >> U_0$), formula (24) has a wider applicability range than formula (22). However, importantly, in addition to providing an expression for the final transition probability in the weak interaction regime, the presented method also accurately treats the temporal dynamics of the molecule formation (Figs. 3 and 4).

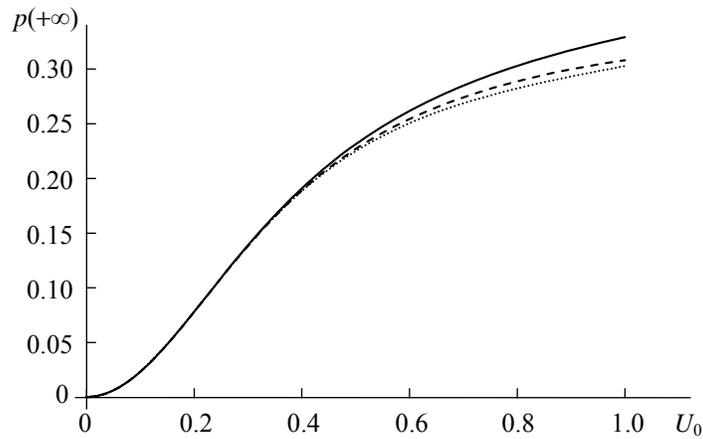

**Fig. 2.** Final transition probability for $\delta_0 = U_0$. Solid line – numerical result, dashed line – approximate formula (22), dotted line – formula (24).

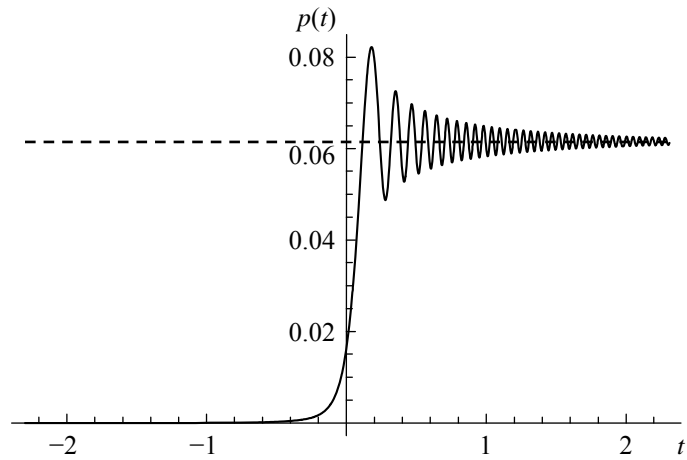

**Fig. 3.** Molecular state probability versus time for $U_0 = 2.5$, $\delta_0 = 70$. For the considered values of the involved parameters, the numerical result and the approximate formula (22) are undistinguishable. Dotted line is the final transition probability given by Eq. (24).

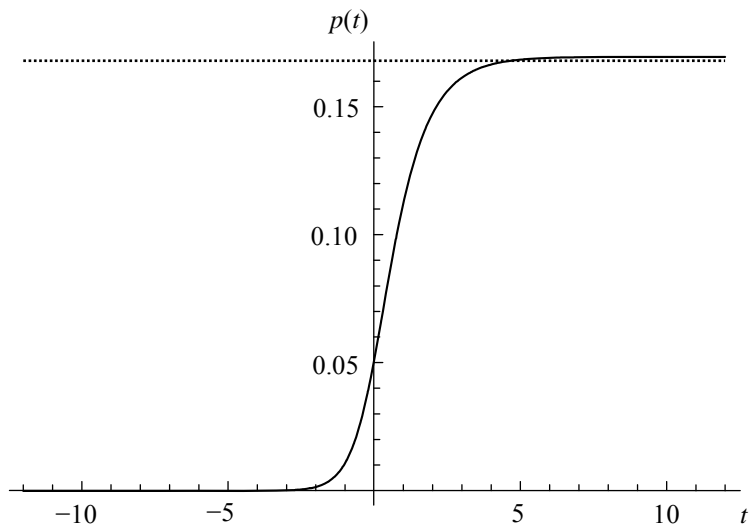

**Fig. 4.** Molecular state probability versus time for $U_0 = 0.3$, $\delta_0 = 0.01$. The numerical solution and the approximate formula (22) produce practically undistinguishable graphs. Dotted line is the final transition probability given by Eq. (24).

## 3. CONCLUSION

We have studied the nonlinear mean-field dynamics of molecule formation at coherent photo- and magneto-association of an atomic Bose–Einstein condensate for the case when the external field configuration is defined by the quasi-linear level-crossing Demkov–Kunike model, characterized by a bell-shaped pulse and finite variation of the detuning, if the photoassociation terminology is used. Using an exact third-order nonlinear differential equation for the molecular state probability, we have constructed a variational ansatz to describe the temporal dynamics of the coupled atom-molecular system in the weak interaction regime, corresponding to the weak coupling limit ($U_0 \ll 1$, $\forall \delta_0$) and the very large detuning regime of the strong coupling limit ($U_0 > 1$, $\delta_0 \gg U_0$). The suggested ansatz is written as a scaled solution to the corresponding linear problem with some effective parameters. Assuming that the parameters involved in the solution of the linear problem are not modified, we have suggested an analytical expression for the scaling parameter. Though in the weak coupling limit the presented formula gives only slightly better prediction for the final transition probability than the one derived in [20] (see Fig. 2), it has an important advantage: it accurately describes the temporal dynamics of the molecule formation in the whole time-domain. Moreover, we have checked numerically that the applicability range of the constructed analytical approximation can considerably be extended if, instead of fixing the fitting parameter $U_0^*$ as $U_0^* = U_0$, we vary it. This would be a notable extension, and we hope to address this question in a future publication.

## ACKNOWLEDGMENTS

This work was supported by the Armenian National Science and Education Fund (ANSEF Grant No. 2009-PS-1692) and the International Science and Technology Center (ISTC Grant No. A-1241). R. Sokhoyan and H. Azizbekyan acknowledge the French Embassy in Yerevan for the Grants No. 2006-4638 and No. 2007-3849 (Boursiers du Gouvernement Français).